\documentclass[epjCONF]{svjour}
\usepackage{graphics}
\usepackage[varg]{txfonts} % Times fonts
\usepackage[latin1]{inputenc}
\session-title{International Conference on New Frontiers in Physics 2012}
\usepackage{textpos}
\begin{document}
\title{Charmonium dissociation and heavy quark transport in hot
   quenched lattice QCD}
\author{
H.-T. Ding\inst{1}\fnmsep\thanks{\email{htding@quark.phy.bnl.gov}} \and 
A. Francis\inst{2}\fnmsep\thanks{\email{francis@kph.uni-mainz.de, speaker}} \and
O. Kaczmarek \inst{3}\fnmsep\thanks{\email{okacz@physik.uni-bielefeld.de}} \and
F. Karsch \inst{1}\fnmsep\inst{3}\fnmsep\thanks{\email{karsch@quark.phy.bnl.gov}}\and
H. Satz \inst{3}\fnmsep\thanks{\email{satz@physik.uni-bielefeld.de}}\and
W. S\"oldner \inst{4}\fnmsep\thanks{\email{wolfgang.soeldner@physik.uni-regensburg.de}} }

\institute{Physics Department, Brookhaven National Laboratory, Upton, New York 11973, USA\and 
Institut f\"ur Kernphysik, Johannes Gutenberg Universit\"at Mainz, D-55099 Mainz, Germany \and 
Fakult\"at f\"ur Physik, Universit\"at Bielefeld, D-33615 Bielefeld, Germany\and
Institut f\"ur Theoretische Physik, Universit\"at Regensburg, D-93040 Regensburg, Germany}

\abstract{
We study the properties of charmonium states at finite temperature in quenched lattice QCD on large
and fine isotropic lattices. We perform a detailed analysis of charmonium correlation and spectral
functions both below and above $T_c$. Our analysis suggests that the S wave states disappear at about 1.5 Tc.
The charm diffusion coefficient is estimated and found to be
approximately $1/\pi T$ at $1.5 T_c \lesssim T \lesssim 3 T_c$.
} %end of abstract
\maketitle
\begin{textblock*}{3cm}(11cm,-10.0cm)
  BI-TP 2012/40.
\end{textblock*}

\section{Introduction}
\label{intro}

In the hot and deconfined QCD medium it has been conjectured that, unlike light mesons, the heavy mesons
might survive and dissociate only at some higher temperature due to Debye screening~\cite{Matsui:1986dk}.
Therefore the suppression of their yield in nucleus-nucleus compared to proton-nucleus or proton-proton collisions may serve as probe for the properties of the medium. 
The experiments carried out at the SPS and LHC at CERN and the RHIC at BNL have indeed observed $J/\psi$ suppression~\cite{Rapp08}. 
The interpretation of experimental data, however, is not as straightforward and in order to disentangle the cold nuclear matter effects on the one hand and hot medium effects on the other, 
it is crucial to have a good understanding of the behavior of heavy quarks and quarkonia in the hot medium.
In addition, a substantial elliptic flow of heavy quarks has been observed~\cite{Adare:2006nq}. As the heavy quark diffusion coefficient $D$ can be related to the ratio of shear viscosity to entropy density $\eta/s$~\cite{Riek:2010py,Moore:20004tg}, a non-perturbative estimate of $D$ is highly desirable.

From the theoretical point of view, the meson spectral function at finite temperature~\cite{Kapusta06} contains all the information on the hadron properties in the thermal medium. This explicitly includes the information
on bound states and transport properties.  As a consequence the non-perturbative computation of spectral functions is of key interest.

In the following we report on a recent success in this direction in quenched lattice QCD. We present the spectral functions in the pseudoscalar and vector channels and estimate the heavy quark diffusion constant.
See \cite{Kaczmarek2012} for a recent overview and \cite{Ding2012} for a more detailed and extended discussion.
 
\section{Euclidean correlation functions and the charmonium spectral function}
\label{sec:theory} 
 
At vanishing momentum charmonium correlation functions are defined via current operators by:
\begin{equation}
G_{H}(\tau,T)=\sum_{\vec{x}} \langle~ J_H(\tau,\vec{x})~J_{H}^{\dag}(0,\vec{0})~\rangle_T .
\end{equation}
 Here $J_H$ is e.g. a local mesonic current operator $\bar{q}(\tau,\vec{x})\Gamma_{H} q(\tau,\vec{x})$, where $\Gamma_{H}=\gamma_{i},\gamma_{5}$ for vector ($V_{ii}$) and pseudo-scalar ($PS$) channels, respectively.
Notice $\langle...\rangle_T$ denotes a thermal average. On the lattice the temperature $T$ is related to Euclidean temporal extent $aN_{\tau}$ by $T=1/(aN_{\tau})$, where $a$ is the lattice spacing.  The above correlation function is exactly related to the spectral function via:
\begin{equation}
G_{H}(\tau,T)=\int_0^{\infty}{\frac{d\omega}{2\pi}~\rho_{H}(\omega,T)}~K(\tau,T,\omega),
\label{eqn:corr}
\end{equation}
where the kernel $K$ is given by $K(\tau,T,\omega)= \mathrm{cosh}(\omega(\tau-\frac{1}{2T}))/\mathrm{sinh}(\frac{\omega}{2T})$.
The spectral function $\rho(\omega)$ contains all the information on the in medium properties of hadrons
and its calculation therefore is our primary goal.
The dissociation temperature can be read off from the deformation of the spectral function, while the heavy quark diffusion constant $D$ is related to the vector spectral function via the Kubo formula:
\begin{equation}
D= \frac{1}{6\chi_{00}}\lim_{\omega\rightarrow0}\sum_{i=1}^{3}\frac{\rho_{ii}(\omega,T)}{\omega},
\label{eqn:ansatz}
\end{equation}
where $\chi_{00}$ is the quark number susceptibility.

To study and estimate the charmonium spectral functions in the pseudoscalar and vector channels as temperature is increased, we first examine the behavior of the correlator itself in a number of ratios and differences. It should be clear that in this fashion we can only obtain a qualitative idea on the properties of the spectral function. The estimates are then used to construct the necessary default models for an analysis based on the maximum entropy method (MEM)~\cite{Asakawa:2000tr}. Using MEM we invert Eq.~\ref{eqn:corr}
and compute the desired spectral function. A detailed analysis of our systematics and statistical uncertainties can be found in ~\cite{Ding2012}. 

\section{Numerical results}
\label{sec:numericalresults}

For our numerical investigation
the standard Wilson plaquette action was used in the gauge sector, while for the (valence) charm quarks we choose the non-perturbatively $\mathcal{O}(a)$ improved Wilson clover fermion action. 
Note, here the mass of vector meson is tuned to be almost the physical $J/\psi$ mass. Employing a fixed scale approach
we  measure correlation functions on very fine quenched lattices sized  $128^{3}\times 96$, $128^3\times48$, $128^3\times32$ and $128^3\times24$. These correspond to the temperatures $0.73~T_c$, $1.46~T_c$, $2.20~T_c$ and $2.93~T_c$, whereby $T_c\simeq270$MeV in quenched QCD. Note, the required
lattice parameters are given in Tab.~\ref{tab:parameters}.

\begin{table}
\centering
\caption{Lattice parameters and number of configurations used in the analysis with a clover improved Wilson fermion action.}
\label{tab:parameters}       % Give a unique label
\begin{tabular}{lllllllll}
\hline\noalign{\smallskip}
$\beta$       & $a$ [fm]  &   $a^{-1}$[GeV]     &  $L_{\sigma}$ [fm]&$c_{\rm SW}$   &   $\kappa$   &      $N_{\sigma}^{3} \times N_\tau$  &      $T/Tc$    & $N_{conf}$  \\
\noalign{\smallskip}\hline\noalign{\smallskip}
7.793           &  0.010     & 18.974                & 1.33                             & 1.310381      &    0.13200    &         $128^{3} \times 96 $                &      0.73      &   234  \\
                    &                &                            &                                     &                      &                      &          $128^{3} \times 48 $                &      1.46        &     461 \\
                    &                &                            &                                     &                      &                      &           $128^{3} \times 32 $                &      2.20      &  105\\
                    &                &                            &                                     &                      &                      &          $128^{3} \times 24 $                &      2.93       & 81 \\
\noalign{\smallskip}\hline
\end{tabular}
\end{table}

\subsection{Charmonium correlation functions}
\label{sec:Corrs}

\begin{figure}
\resizebox{0.5\columnwidth}{!}{\includegraphics{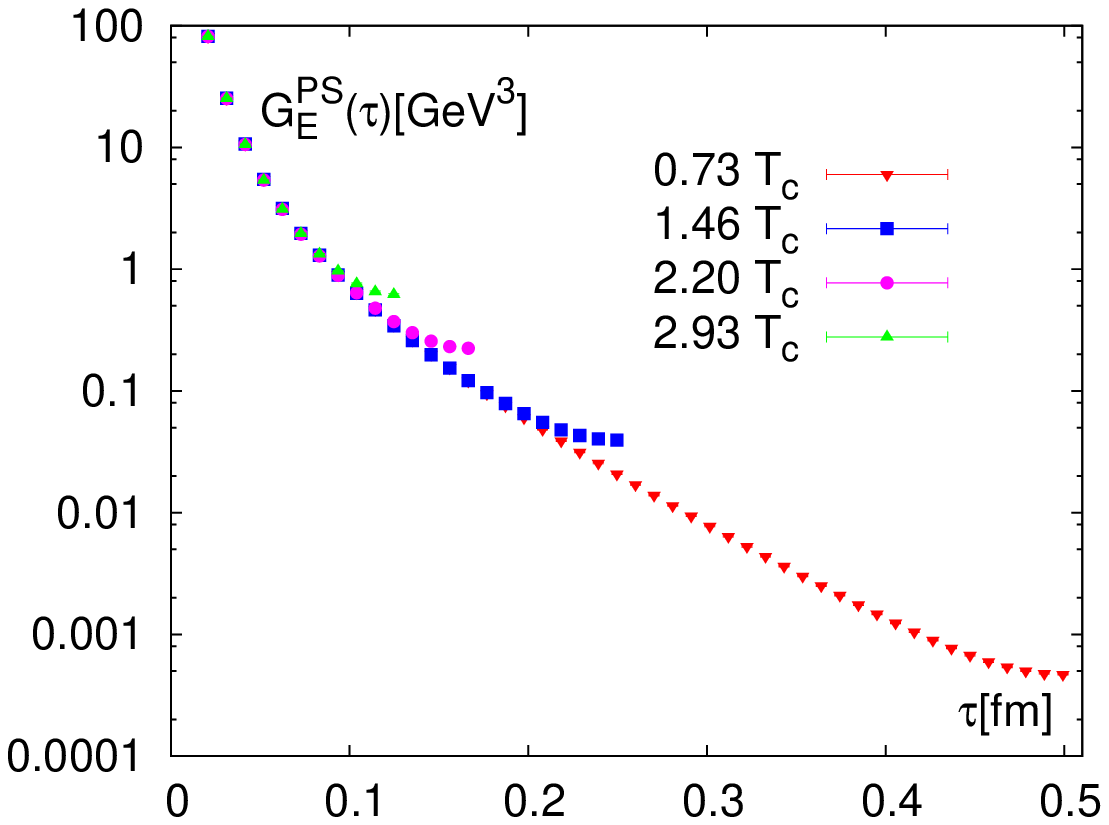} }
\resizebox{0.5\columnwidth}{!}{\includegraphics{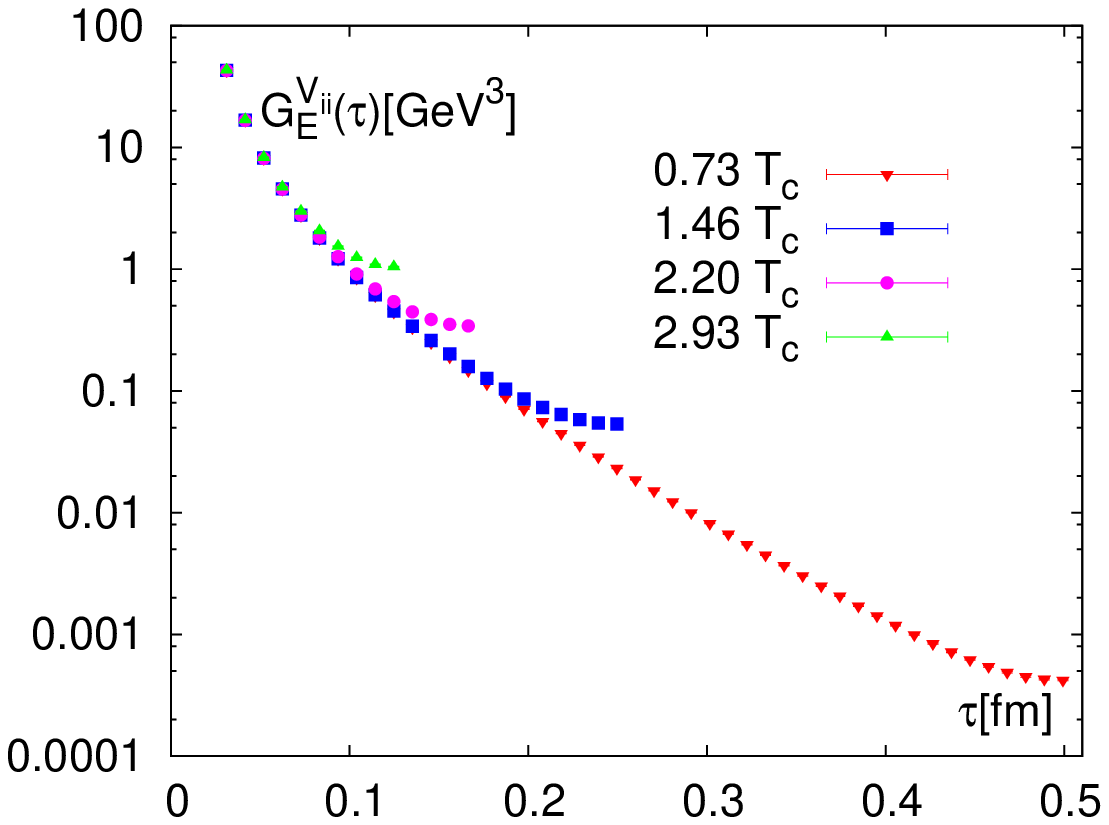} }
\caption{Euclidean correlation functions $G(\tau)[\textrm{GeV}^3]$ in the pseudoscalar $PS$(left) and vector $V_{ii}$(right) channels at $\beta=7.793$ ($a=0.010$ fm) and the temperatures $T=0.73$, 1.46, 2.20 and 2.93$~T_c$ }
\label{fig:Corr}       % Give a unique label
\end{figure}

To begin our analysis we show the correlation functions in the pseudoscalar and vector channels in physical units in Fig.~\ref{fig:Corr}.  Clearly all correlators agree throughout most of the available Euclidean time distance except when they near their respective midpoints. This small effect in the large distance region is the essential piece for understanding the thermal modification of the bound state and transport regions of the spectral function, while the correlation function at small to intermediate distances is dominated by the continuum contribution. As a consequence a more sophisticated analysis highlighting the thermal modifications by taking e.g. appropriate ratios or differences is necessary.

To analyze the properties of the underlying spectral functions on the correlator level more thoroughly we turn
to studying it in conjuncture with the so called reconstructed correlator. \\
In a reconstructed correlator the temperature dependence of the kernel in Eq.~\ref{eqn:corr} is taken into account, while the spectral function remains unchanged. It is particularly interesting to compare a zero or low temperature spectral function with a thermal one and as it turns out in this case it is possible to calculate the reconstructed correlator analytically as follows, first one constructs~\cite{Datta04}:
\begin{equation}
G_{rec}(\tau,T;T^\prime) = \int_0^{\infty} \frac{d\omega}{2\pi}~\rho(\omega,T^\prime)\,\,\frac{\cosh\left(\omega(\tau - 1/2T)\right)}{\sinh(\omega/2T)},
\label{eqn:rec_cor}
\end{equation}

At this point one would generally need the spectral function $\rho(\omega,T^{\prime})$ at a reference temperature $T^{\prime}$ and consequently the evaluation of
Eq.~(\ref{eqn:rec_cor}) would suffer from the uncertainty of the determination of that spectral function. However
there is a useful and exact relation, which is a generalization of Ref.~\cite{Meyer10}:
\begin{equation}
\frac{\cosh[\omega(\tau-N_{\tau}/2)]}{\sinh (\omega N_{\tau}/2)}~~\equiv~\sum_{\tau^{\prime}=\tau;~\tau^{\prime}+=N_{\tau}}^{N_{\tau}^{\prime}-N_{\tau}+\tau} \frac{\cosh[\omega(\tau^{\prime}-N_{\tau}^{\prime}/2)]}{\sinh (\omega N_{\tau}^{\prime}/2)},
\label{eqn:kernel_rules}
\end{equation}
where $T^{\prime}=(a N_{\tau}^{\prime})^{-1},~~T=(aN_{\tau})^{-1},~~\tau^{\prime}\in[0,~N_{\tau}^{\prime}-1],~~\tau\in[0,~N_{\tau}-1],~~N_{\tau}^{\prime}=m~ N_{\tau},~~m\in\mathbb{Z}^{+}$. $N_{\tau}$ and $N_{\tau}^{\prime}$ are the number of time slices in the temporal directions at temperature $T$ and $T^{\prime}$, respectively. $\tau$ denotes the time slice of the correlation function at temperature $T$ while $\tau^{\prime}$ denotes the time slice of the correlation function at 
temperature $T^{\prime}$. The sum of $\tau^{\prime}$ on the right hand side of  Eq.~(\ref{eqn:kernel_rules}) starts from $\tau^{\prime}=\tau$ with a step length $N_{\tau}$ to the upper limit $N_{\tau}^{\prime}-N_{\tau}+\tau$. After putting $\rho(\omega,T^{\prime})$ into both sides of the above relation and performing the integration over $\omega$, one immediately arrives at \cite{Ding2012}:
\begin{equation}
G_{rec}(\tau,T;T^{\prime}) = \sum_{\tau^{\prime}=\tau;~\Delta\tau^{\prime}=N_{\tau}}^{N_{\tau}^{\prime}-N_{\tau}+\tau}  G(\tau^{\prime},T^{\prime}),
\label{eqn:Grec_data}
\end{equation}
with this relation we are therefore capable of computing the reconstructed correlator directly from our low temperature results at $N_\tau=96$ without having to extract the spectral function, thereby bypassing a systematic source of error in this quantity.

\begin{figure}
\resizebox{0.5\columnwidth}{!}{\includegraphics{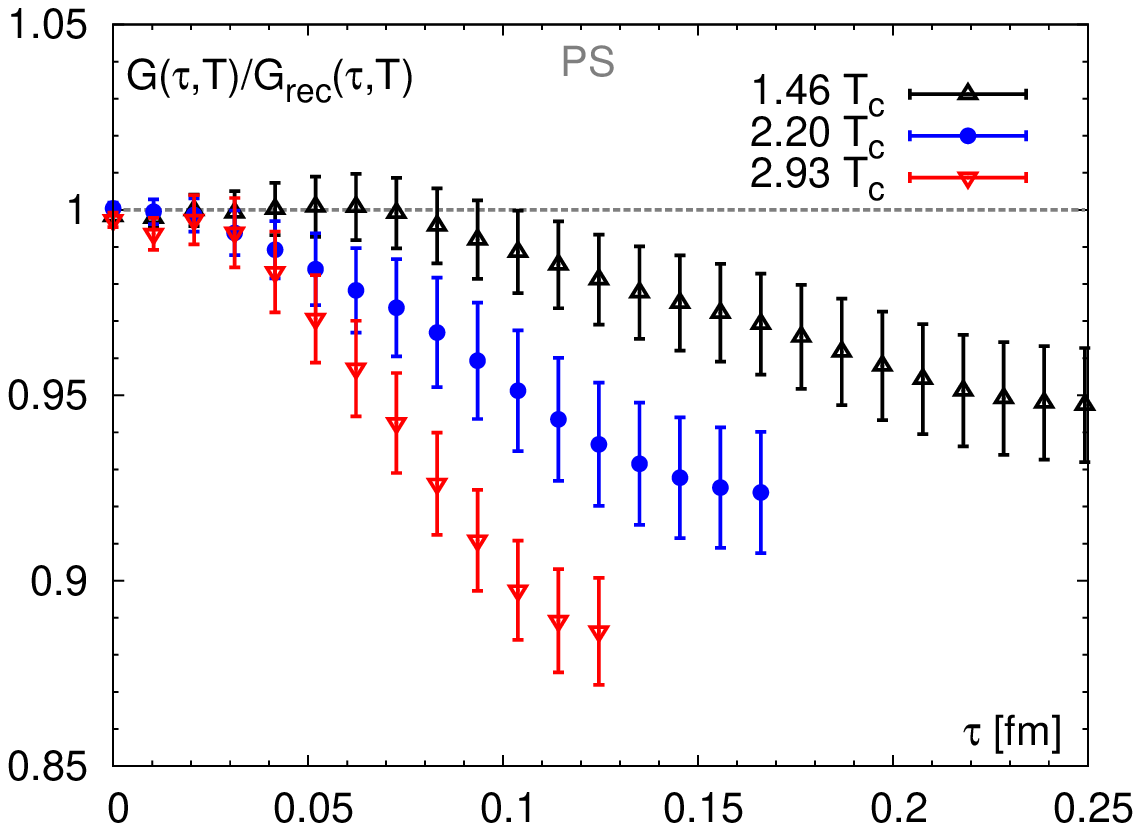} }
\resizebox{0.5\columnwidth}{!}{\includegraphics{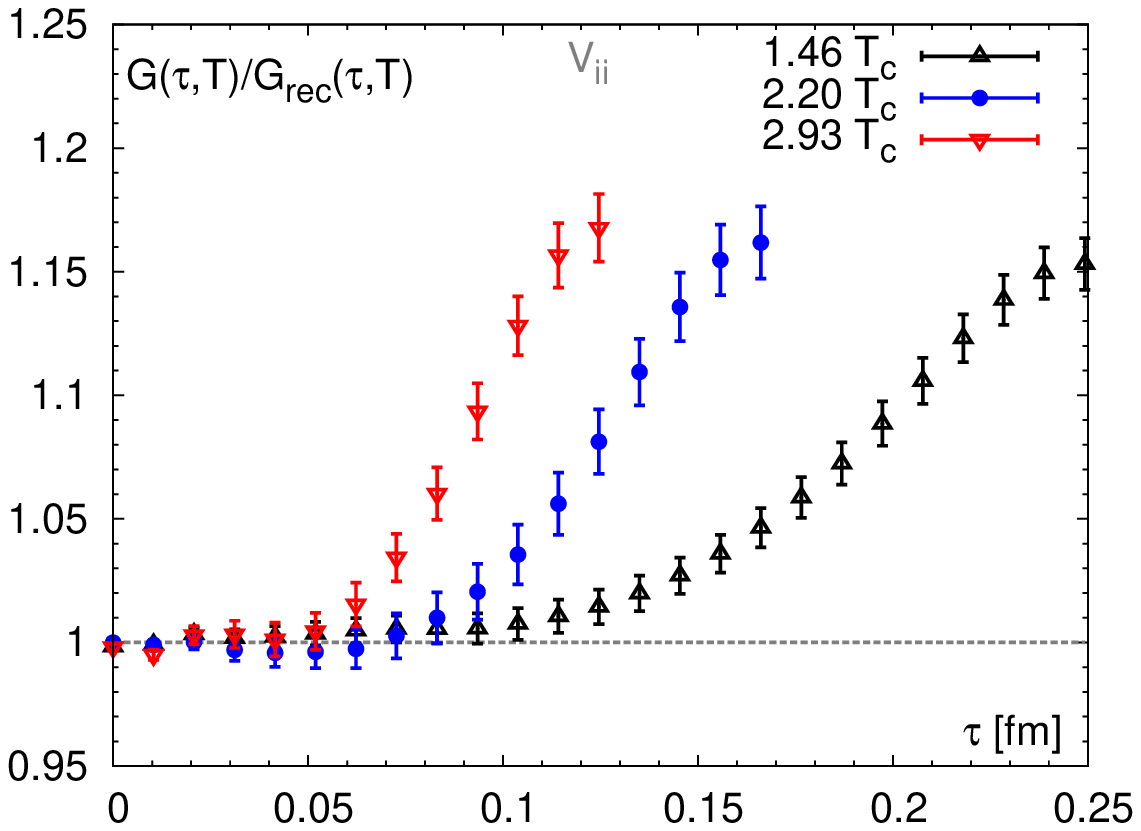} }
\caption{The ratio $G(\tau,T)/G_{rec}(\tau,T)$ for $PS$ (left) and $V_{ii}$ (right) channels as a function of the Euclidean distance $\tau$
at $T=1.46$, 2.20 and 2.93$~T_c$. The reconstructed correlator $G_{rec}$ is obtained directly from correlator data at $0.73~T_c$.}
\label{fig:GGrec}       % Give a unique label
\end{figure}
Given this insight we use Eq.~\ref{eqn:Grec_data} to compute the ratio of the thermal and reconstructed correlators $G(\tau)/G_{rec}(\tau)$ at $1.46T_c$, $2.20T_c$ and $2.93T_c$ in the pseudoscalar and vector channels, where the results are shown in Fig.~\ref{fig:GGrec}.
Concentrating first on the pseudoscalar case in Fig.~\ref{fig:GGrec}(left) we observe that the ratios decrease monotonically at the available temperatures as the Euclidean time separation increases. Note, at the largest distances, the ratios deviate from unity by about 5\%, 8\% and 12\% at 1.46, 2.20 and 2.93 $T_c$, respectively. 
Turning to the right of Fig.~\ref{fig:GGrec} and therefore the $V_{ii}$ channel we see an entirely different behavior. Here the ratios increase monotonically with increasing distance, reaching a deviation of about 16\%
at the large distances.
These results are interesting as there is no emerging transport contribution in the $PS$ channel, therefore our results hint at the decrease in this channel being due to bound state modification, while a large part of the modification in the $V_{ii}$ channels is due to the emerging transport contribution.

%To study this more closely we define two ratios, which are approximately independent of a zero mode contribution \cite{Petreczky08}, first:
%\begin{equation}
%\frac{G^{\rm diff}(\tau,T)}{G^{\rm diff}_{\rm rec}(\tau,T)} \equiv \frac{G(\tau,T) -  G(\tau+1,T)}{G_{\rm rec}(\tau,T) -  G_{\rm rec}(\tau+1,T)},
%\label{eq:GoverGrec_diff}
%\end{equation}
%which equals the ratio of the time derivative of the correlators to the time derivative of the reconstructed correlators at $\tau+1/2$. And second:
%\begin{equation}
%\frac{G^{\rm sub}(\tau,T)}{G^{\rm sub}_{\rm rec}(\tau,T)} \equiv \frac{G(\tau,T) -  G(N_{\tau}/2,T)}{G_{\rm rec}(\tau,T) -  G_{\rm rec}(N_{\tau}/2,T)},
%\label{eq:GoverGrec_midpoint-subtracted}
%\end{equation}
%which is the ratio of midpoint subtracted correlators~\cite{Umeda07}.
%
To study this more closely we define two ratios, which are approximately independent of a zero mode contribution:
\begin{equation}
\frac{G^{\rm diff}(\tau,T)}{G^{\rm diff}_{\rm rec}(\tau,T)} \equiv \frac{G(\tau,T) -  G(\tau+1,T)}{G_{\rm rec}(\tau,T) -  G_{\rm rec}(\tau+1,T)},\quad\textrm{and}\quad
\frac{G^{\rm sub}(\tau,T)}{G^{\rm sub}_{\rm rec}(\tau,T)} \equiv \frac{G(\tau,T) -  G(N_{\tau}/2,T)}{G_{\rm rec}(\tau,T) -  G_{\rm rec}(N_{\tau}/2,T)}.
\label{eq:GoverGrec_midpoint-subtracted}
\end{equation}
The first equals the ratio of the time derivative of the correlators to the time derivative of the reconstructed correlators at $\tau+1/2$~\cite{Petreczky08}, while the second is the ratio of midpoint subtracted correlators~\cite{Umeda07}.\\
\begin{figure}
\resizebox{0.5\columnwidth}{!}{\includegraphics{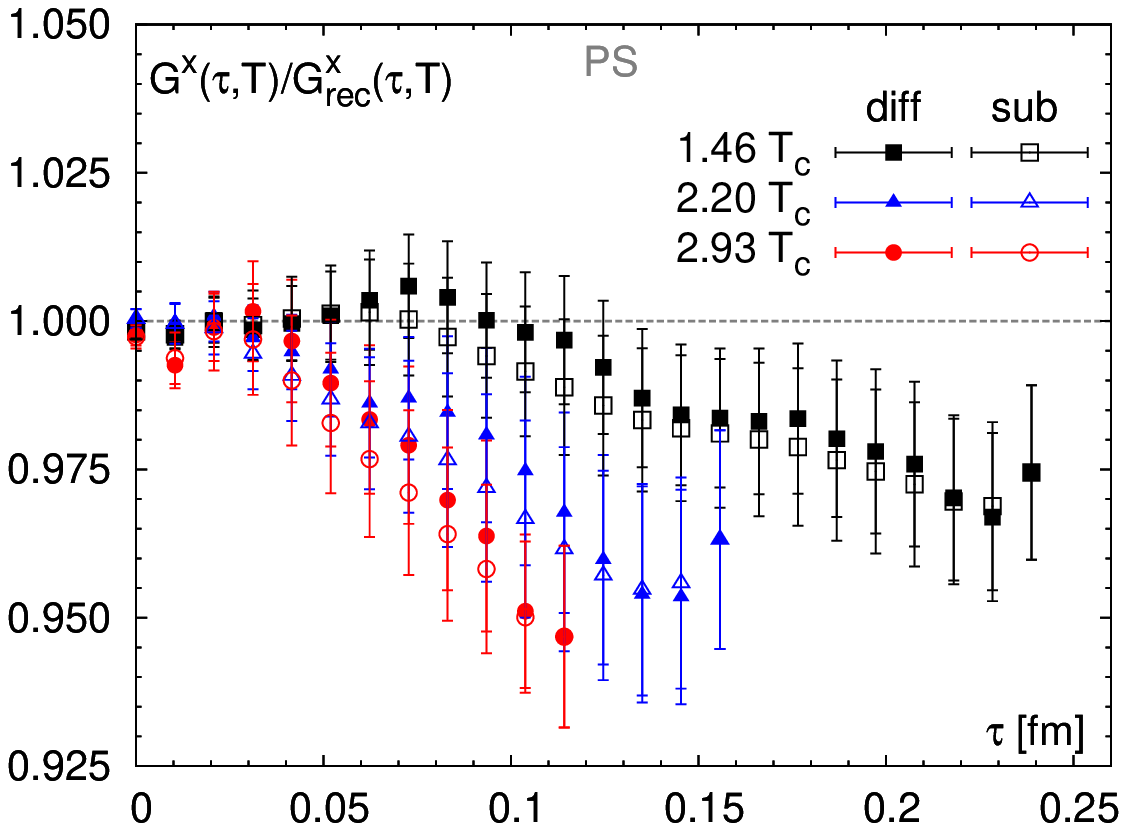} }
\resizebox{0.5\columnwidth}{!}{\includegraphics{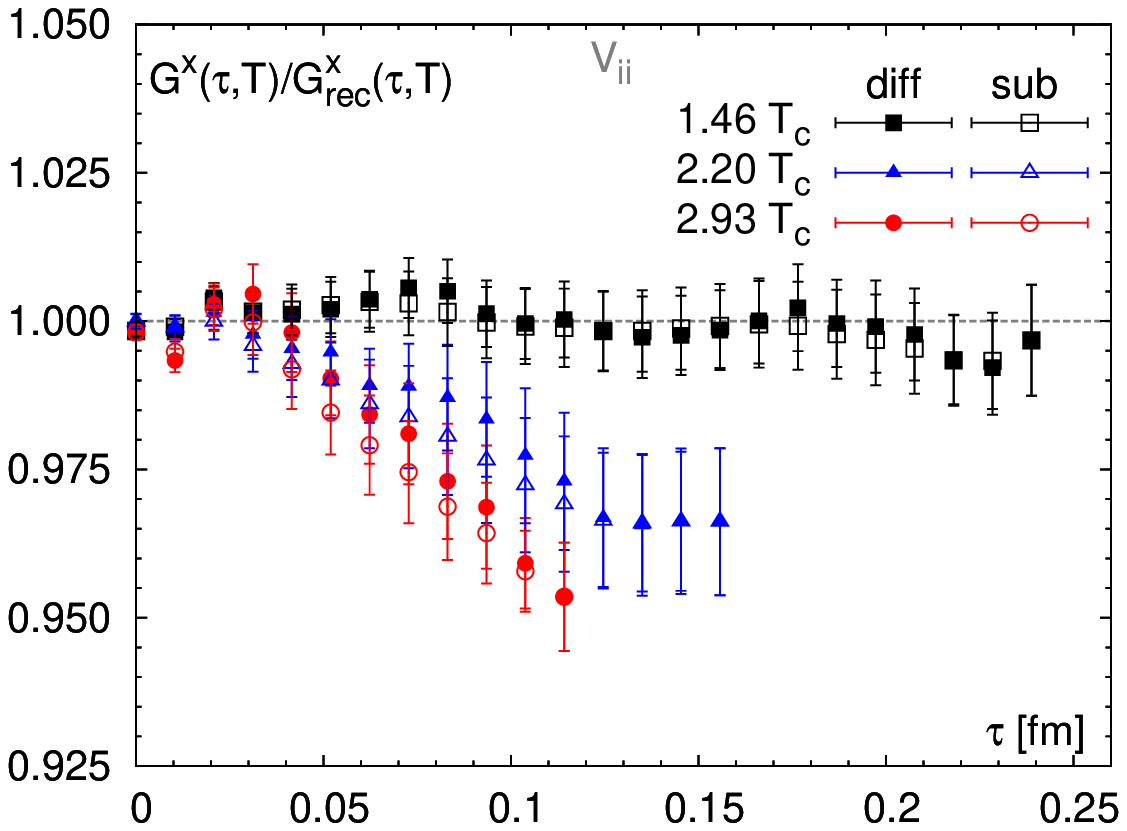} }
\caption{The ratio $G^{\rm diff}(\tau,T)/G^{\rm diff}_{rec}(\tau,T)$ ($G^{\rm sub}(\tau,T)/G^{\rm sub}_{rec}(\tau,T)$) for the $PS$(left) and $V_{ii}$(right) channels as a function of the Euclidean distance $\tau$  at $T=1.46$, 2.20 and 2.93$~T_c$. The superscript ``x" denotes either ``diff" or ``sub". }
\label{fig:Gdiff}       % Give a unique label
\end{figure}
As noted above both ratios are approximately independent of a zero mode contribution, consequently they should be almost independent of transport phenomena. The corresponding results in the $PS$ channel are given in Fig.~\ref{fig:Gdiff}(left) and we find that both $G^{diff}(\tau)/G^{diff}_{rec}(\tau)$ and $G^{sub}(\tau)/G^{sub}_{rec}(\tau)$ give very similar results. Throughout the available temperature range we retain
the decreasing trend with increasing distance from the analysis of $G(\tau)/G_{rec}(\tau)$, although the magnitude of the deviation has decreased, the values at the largest distance being shifted up by about 3\% at both $1.46~T_c$ and $2.20~T_c$ and about 6\% at $2.93~T_c$. As there is no transport contribution in this channel we would have expected fairly small changes from the behavior of $G(\tau)/G_{rec}(\tau)$ to $G^{diff}(\tau)/G^{diff}_{rec}(\tau)$ or $G^{sub}(\tau)/G^{sub}_{rec}(\tau)$ and indeed we see this confirmed.\\
Turning to the $V_{ii}$ channel on the right of Fig.~\ref{fig:Gdiff} we first note that $G^{diff}(\tau)/G^{diff}_{rec}(\tau)$ and $G^{sub}(\tau)/G^{sub}_{rec}(\tau)$ give similar results as in the $PS$ case. The behavior compared to $G(\tau)/G_{rec}(\tau)$ on the other hand has been greatly modified. As such the data for $1.46T_c$ instead of increasing now is almost flat, while $2.20T_c$ and $2.93T_c$ show a decreasing trend similar to the $PS$ case. From this we infer that most of the temperature dependence in $G(\tau)/G_{rec}(\tau)$ in the $V_{ii}$ channel is in fact due to the transport contribution.\\

\begin{figure}
\resizebox{0.5\columnwidth}{!}{\includegraphics{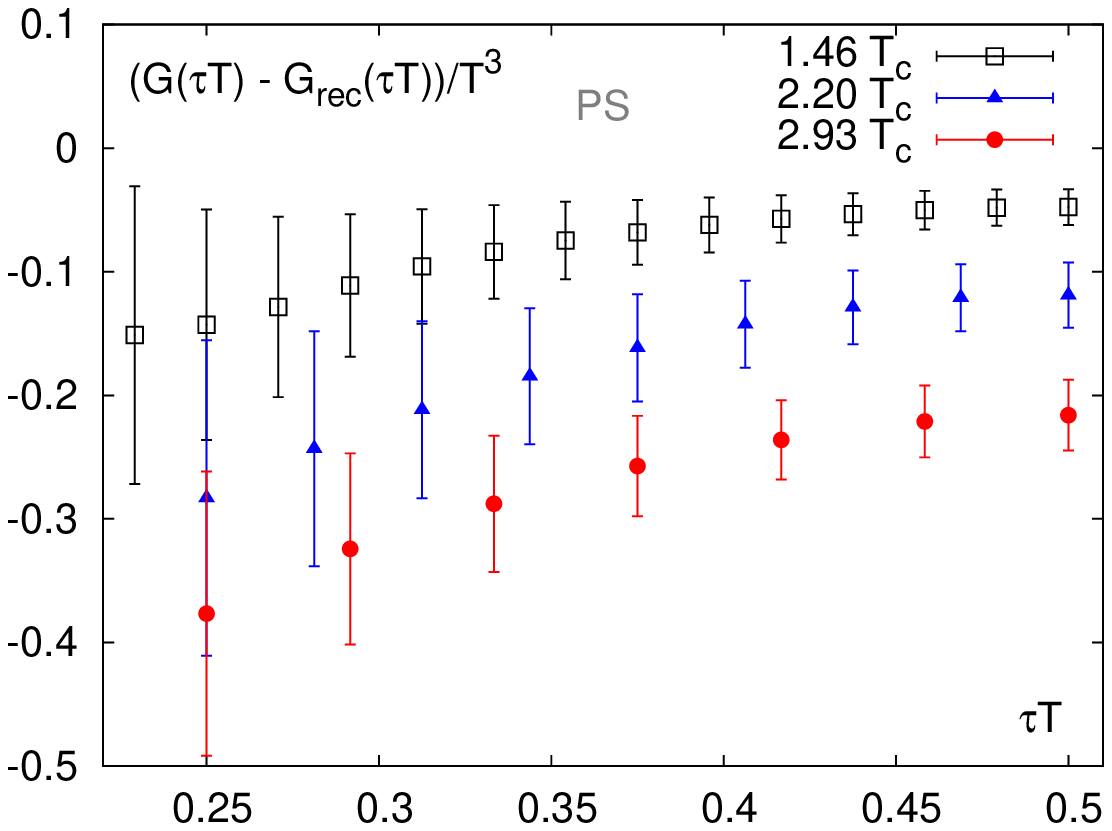} }
\resizebox{0.5\columnwidth}{!}{\includegraphics{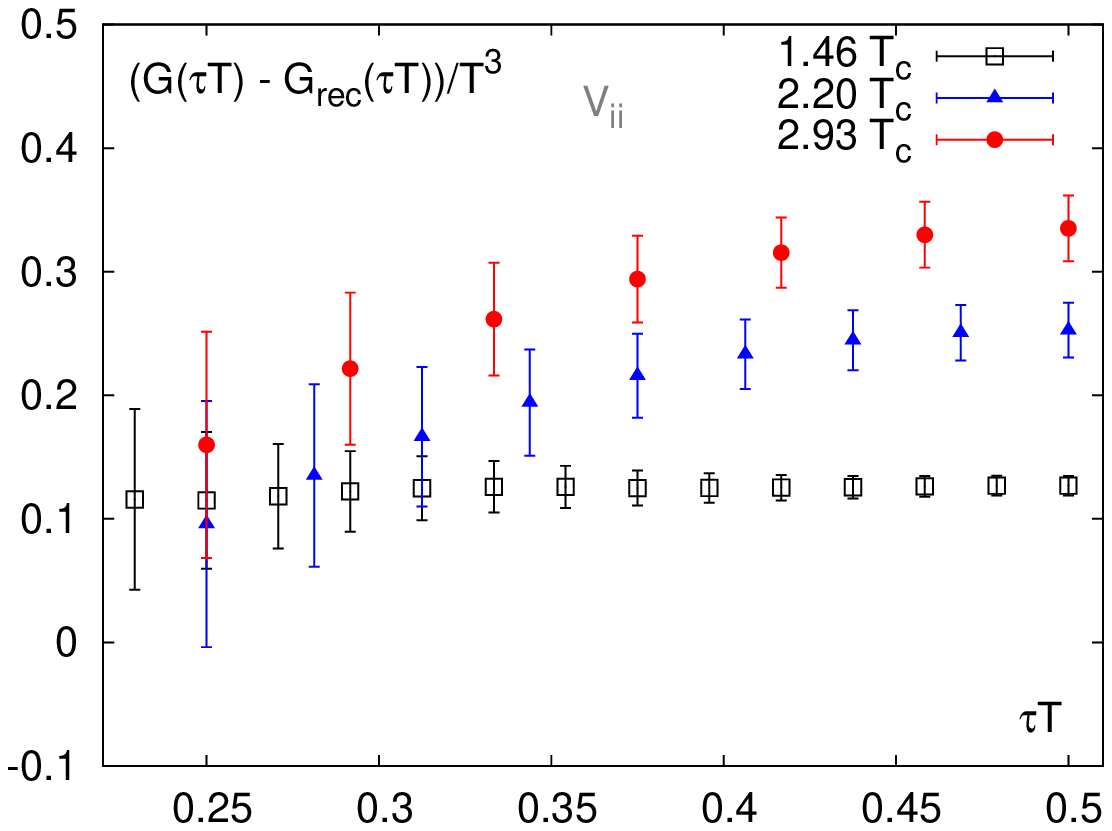} }
\caption{ $(G(\tau)-G_{rec}(\tau))/T^3$  in the $PS$(left) and $V_{ii}$(right) channels as a function of the Euclidean distance $\tau$ at $T=1.46$, 2.20 and 2.93$~T_c$.}
\label{fig:G-Grec}       % Give a unique label
\end{figure}

Another way to look at the data is to take the difference $G(\tau)-G_{rec}(\tau)$, implying:
\begin{equation}
G(\tau)-G_{rec}(\tau) = \int_0^{\infty} \frac{d\omega}{2\pi}~\Big[\rho(\omega,T)-\rho(\omega,T^\prime)\Big]\,\,\frac{\cosh\left(\omega(\tau - 1/2T)\right)}{\sinh(\omega/2T)},
\end{equation}
consequently this difference between the measured correlator and the reconstructed correlator provides information on the difference of spectral functions below and above $T_c$ and the corresponding results
are given in Fig.~\ref{fig:G-Grec}.
Note in both $V_{ii}$ and  $PS$ channels we observe that $G(\tau) - G_{rec}(\tau )$ increases with $\tau T$ at the available temperatures. In the $PS$ case on the left of Fig.~\ref{fig:G-Grec} however all values are negative, while for the $V_{ii}$ on the right they are positive.\\
As $G(\tau) - G_{rec}(\tau )$ probes the difference of the two spectral functions $\Delta\rho(\omega)$ the results in the $PS$ channel require $\Delta\rho(\omega)<0$ in a certain $\omega$ region. With no transport contribution in this channel and given the positivity condition of the spectral functions this implies the low temperature bound state peak overcompensates any remnants of the peak at finite temperature.\\
At the same time in the $V_{ii}$ channel the positive values indicate a more intricate situation, as the emergence of transport leads to a positive contribution, while an overcompensation of the bound state would lead to a negative part.\\
Note, the weakest $\tau$-dependence is seen in the $V_{ii}$ channel at $1.46T_c$. Here we observe an almost flat behavior with positive values throughout. Assuming that there is a complete cancellation of bound state peaks at this temperature in $\Delta\rho(\omega)$ we may try to fit the transport contribution to the
Ansatz defined in Eq.~\ref{eqn:ansatz}. In this case there is only one free parameter and the uncertainty in the heavy quark mass, for $M=1.0$GeV and $M=1.8$GeV we obtain:
\begin{equation}
M=1.0~{\rm GeV}, ~~ 2\pi T D \approx 0.6 \quad{\rm and}\quad
M=1.8 ~{\rm GeV},~~ 2\pi T D \approx 3.6  .
\end{equation}
Even though it is not convincing to use a one parameter fit here to obtain a robust value for the heavy quark diffusion coefficient, these results are in the right ballpark and provide useful input for our subsequent MEM analysis.

\subsection{Charmonium spectral functions via MEM}
\label{sec:spfs}

\begin{figure}
\resizebox{0.5\columnwidth}{!}{\includegraphics{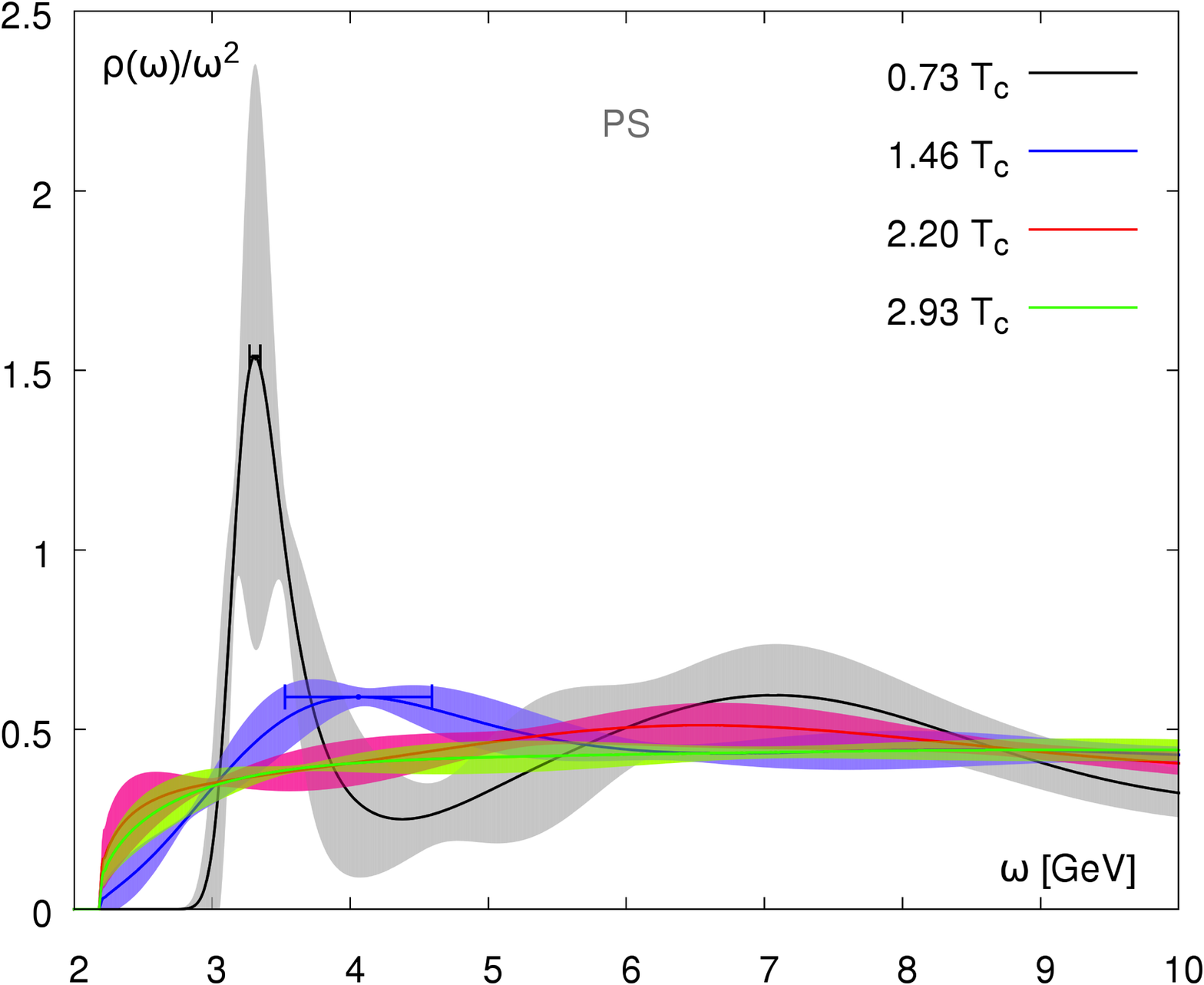} }
\resizebox{0.5\columnwidth}{!}{\includegraphics{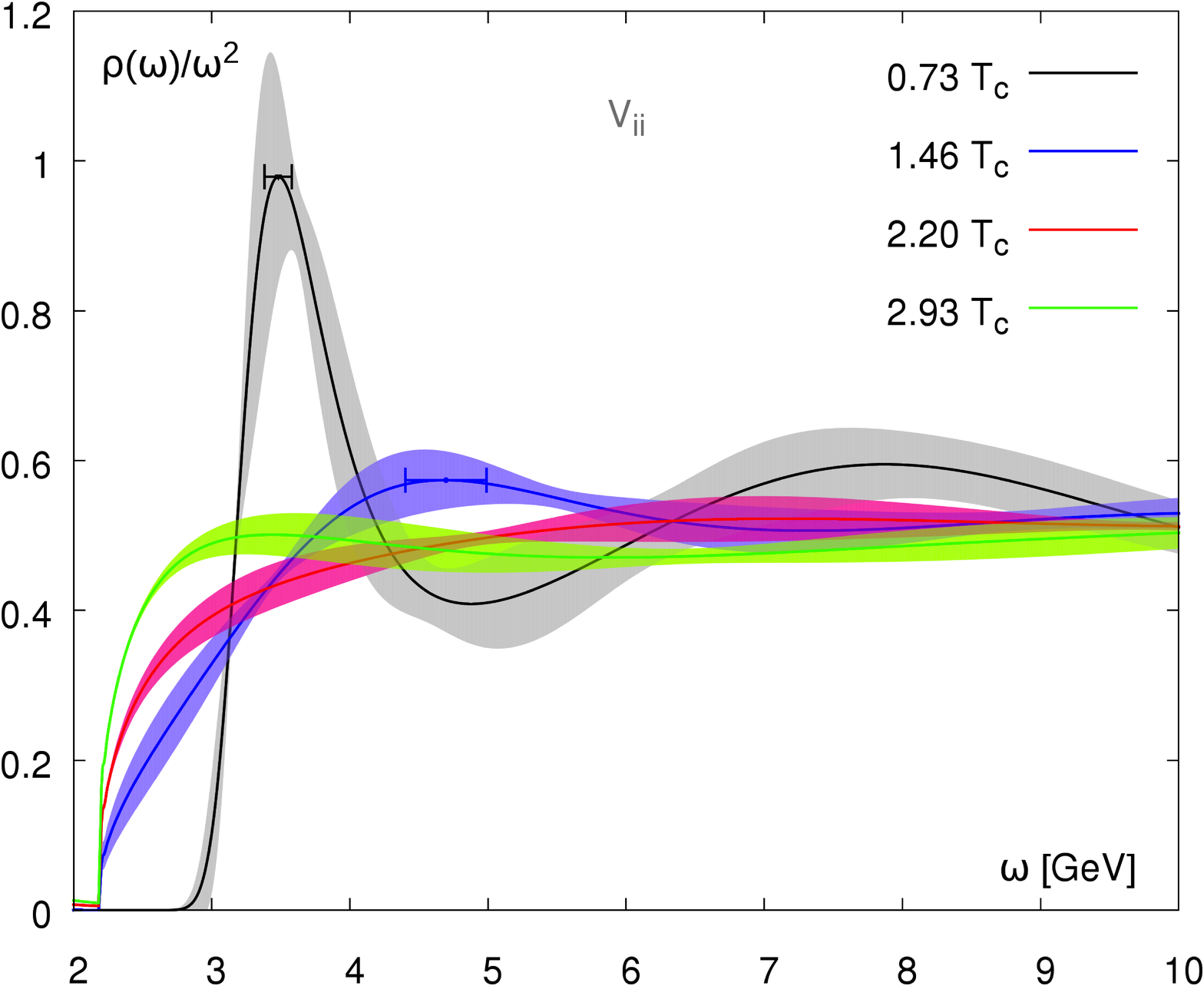} }
\caption{Statistical uncertainties of output spectral functions in $PS$ (left) and $V_{ii}$ (right)
channels at all available temperatures. The shaded areas are statistical errors of amplitudes of output spectral
functions from Jackknife analyses and the solid lines inside the shaded areas are mean values
of spectral functions. The horizontal error bars at the first peaks of spectral functions at $0.73~T_c$ an $1.46~T_c$ stand for the statistical uncertainties of the peak location obtained from Jackknife analyses.}
\label{fig:spf}       % Give a unique label
\end{figure}

Clearly the comparison of the measured correlator with the reconstructed correlator can only give a rough idea of the magnitude of any medium effects at a certain temperature. To really explore the properties of the thermal charmonium states, one has to go to the level of spectral functions.\\
Here we do this by using the maximum entropy method (MEM) to extract the spectral functions directly from the correlator data.
In the MEM analysis presented here, we use as number of frequencies $N_{\omega}=8000$, the minimum energy $a\omega_{min}=0.000001$ and implement the modified kernel $K^{\ast}=\mathrm{tanh}(\omega/2)\cdot K$~\cite{Ding09_2,Engels10} in order to explore the low energy behavior of spectral function~\cite{Aarts07}. 
The default model in the $PS$ channel is a normalized free lattice spectral function, while in the $V_{ii}$ channel it is a normalized free lattice spectral function with a transport peak based on Eq.~\ref{eqn:ansatz}
and the diffusion coefficient estimated above.\\
The results of this analysis are given in Fig.~\ref{fig:spf}, where we show the $PS$(left) and $V_{ii}$(right) spectral functions divided by $\omega^2$ at the available temperatures above $T_c$ and also below $T_c$.
Note, throughout we show the spectral functions in $PS$ and $V_{ii}$ channels with error bands given by the statistical uncertainties, while the error of the peak positions, see \cite{Ding2012}, are given as horizontal error bars.

In Fig.~\ref{fig:spf}(left) we observe that at $0.73~T_c$ the spectral function in the $PS$ channel has large uncertainties in the amplitude at the point which corresponds to the ground state peak location in the mean spectral function. However, even at the lower end of the error band, the amplitude is still larger than the peak amplitudes at the higher temperatures within the errors.\\
Unlike the large uncertainties shown in the amplitude of the peak height, the peak location of the ground state peak at $0.73~T_c$ is well determined. A Jackknife analysis yields $m_{\eta_c}=3.31 (4)$ GeV. At $1.46~T_c$ this peak is shifted by about 0.8 GeV to around 4.1 GeV. At $2.23~T_c$ there is hardly a peak structure that can be identified within the statistical uncertainties. At $2.93~T_c$ the spectral function flattens further. Thus this picture suggests that $\eta_c$ is melted already at  $1.46~T_c$.

Turning to the right of Fig.~\ref{fig:spf} we analyze the resonance part of the spectral function in the $V_{ii}$ channel. We observe that the peak location of the spectral function at $0.73~T_c$ does not have an overlap with the peak location of the spectral function at $1.46~T_c$ and the amplitudes between these two differ a lot (see horizontal error bars). At both $2.20~T_c$ and $2.93~T_c$ there is hardly any peak structure. As such this picture indicates that also $J/\psi$ is already dissociated at $1.46~T_c$.

\begin{figure}
\resizebox{0.5\columnwidth}{!}{\includegraphics{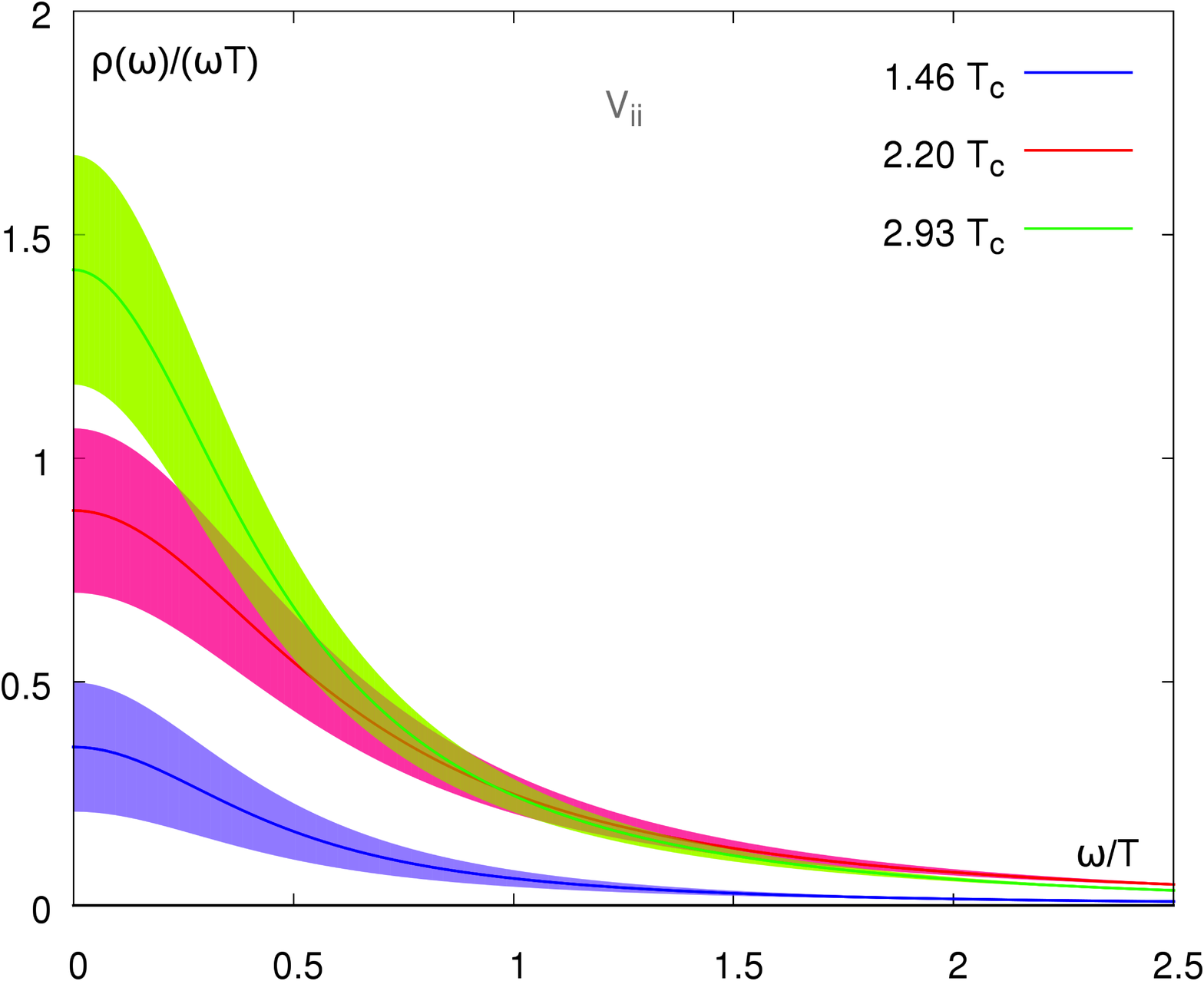} }
\resizebox{0.5\columnwidth}{!}{\includegraphics{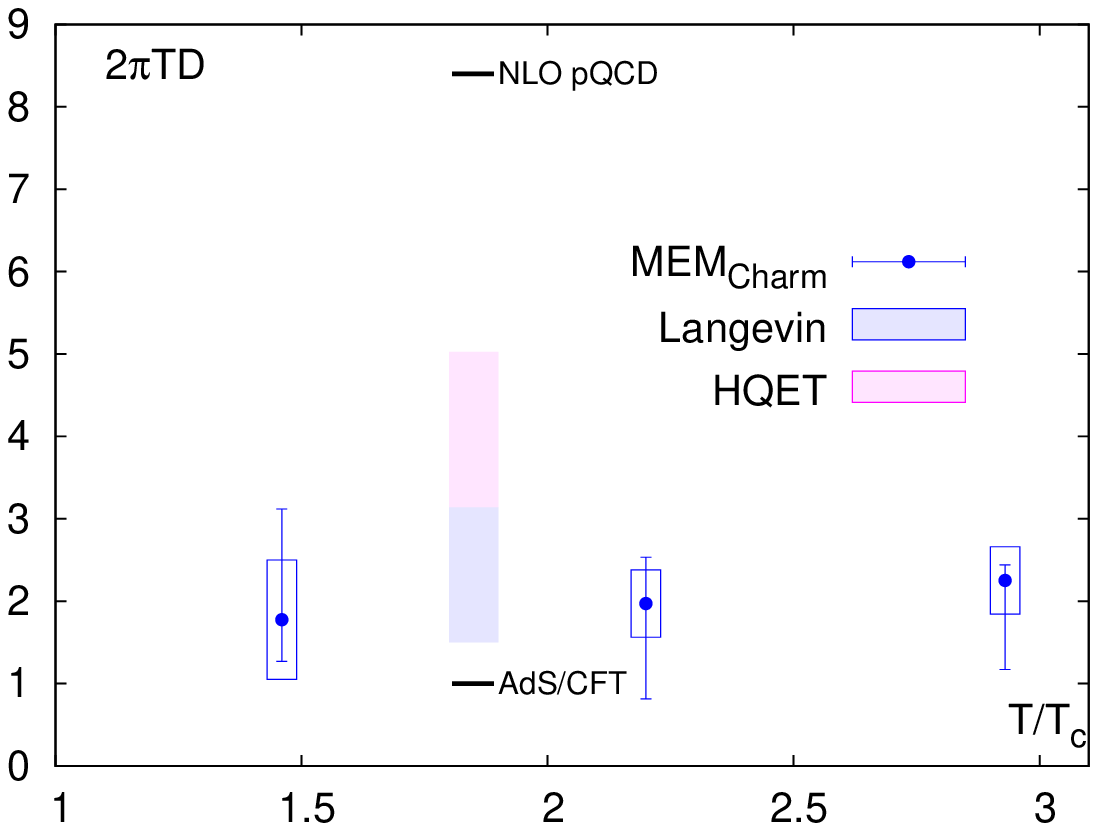} }
\caption{Left: statistical uncertainties of transport peaks at $T>T_c$. Right: the resulting charm diffusion coefficients. The boxes stand for  statistical error estimated from Jackknife method while the bars stand for  systematic uncertainties from MEM analyses. To compare also the results from perturbation theory~\cite{CaronHuot:2007gq}, Langevin dynamics~\cite{Moore:20004tg}, Ads/CFT~\cite{Kovtun:2003wp} and HQET \cite{Banerjee:2011ra,Francis:2011gc} are given.}
\label{fig:diffusion}       % Give a unique label
\end{figure}
Focusing on the very low frequency part of the spectral function given in the vector channel, we show the spectral function divided by $\omega T$ in Fig.~\ref{fig:diffusion}(left).
Notice here the statistical uncertainties on the amplitude of the peak are relatively small. The charm diffusion coefficient is related to the amplitude of the transport peak at vanishing frequency through the Kubo formula (Eq.~\ref{eqn:ansatz}). The corresponding estimates for the heavy diffusion coefficient $D$ are summarized in the right plot of Fig.~\ref{fig:diffusion}. The boxes denote the statistical uncertainties and the error bars reflect systematic uncertainties obtained from the analyses discussed in \cite{Ding2012}.
Note the charm diffusion coefficient obtained at $1.46~T_c$ is the most reliable one among the three temperatures above $T_c$ since more prior information is known at this temperature. At $2.20~T_c$ and $2.93~T_c$, due to the lack of precise prior information and also less number of data points that can be used in the MEM analyses, the uncertainties on the charm diffusion coefficient thus might be underestimated.

For comparison we highlight results from alternative approaches to the calculation of the heavy quark diffusion coefficient in Fig.~\ref{fig:diffusion}(right), showing the perturbative next-to-leading
order results of ~\cite{CaronHuot:2007gq}, those from a computation in AdS/CFT \cite{Kovtun:2003wp}, as well as those from Langevin dynamics derived in \cite{Moore:20004tg}.
Additionally we show results from a recently proposed method to compute the momentum diffusion coefficient $\kappa$ from a purely gluonic operator derived from heavy quark effective theory (HQET)~\cite{CasalderreySolana:2006rq,CaronHuot:2009uh}.
Hereby $\kappa$ is related to the heavy quark diffusion coefficient by $\kappa=2T^2/D$. First calculations from two groups on this operator agree fairly well~\cite{Banerjee:2011ra,Francis:2011gc} and  compare favourably with our results. %Note, on this qualitative level all results are compatible with estimates derived from the heavy flavor $R_{AA}$ and $v_w$ PHENIX data \cite{Adare:2006nq}.

\section{Summary}
In summary our analysis of charmonium correlation functions on large isotropic quenched gauge configurations using Wilson clover fermions implies that both $J/\psi$ and $\eta_c$ have dissociated by a temperature of $1.46T_c$. At the same time we estimated the charmonium diffusion coefficient and found a value $2\pi T D\simeq 1.8-2.3$ in the available temperature range.

\section*{Acknowledgments}
\label{ackn}
This work has been supported in part by the Deutsche Forschungsgemeinschaft under grant GRK 881 and by contract DE-AC02-98CH10886 with the U.S. Department of Energy. 
Numerical simulations have been performed on the BlueGene/P at the New York Center for Computational Sciences (NYCCS) which is supported by the State of New York and 
the BlueGene/P at the John von Neumann Supercomputer Center (NIC) at FZ-J\"ulich, Germany.

%\begin{figure}
%\resizebox{0.75\columnwidth}{!}{%
%  \includegraphics{fig1.eps} }
%\caption{Please write your figure caption here.}
%\end{figure}
%\begin{table}
%\caption{Please write your table caption here.}
%\begin{tabular}{lll}
%\hline\noalign{\smallskip}
%\end{tabular}
%\end{table}
%
%\begin{thebibliography}{}
%Author, Journal \textbf{Volume}, (year) page numbers
%Author, \textit{Book title} (Publisher, place year) page numbers
% etc
%\end{thebibliography}

\end{document}